\documentclass[aps,prb,twocolumn,showpacs,groupedaddress,preprintnumbers,amsmath,amssymb]{revtex4}

\usepackage{epsfig}
\usepackage{graphicx}
\usepackage{dcolumn}
\usepackage{bm}

\begin{document}

\title{Coulomb correlation effects in LaOFeAs: LDA+DMFT(QMC) study}

\author{A.O.~Shorikov}
\affiliation{Institute of Metal Physics, Russian Academy of Sciences,
620041 Yekaterinburg GSP-170, Russia}

\author{M.A.~Korotin}
\affiliation{Institute of Metal Physics, Russian Academy of Sciences,
620041 Yekaterinburg GSP-170, Russia}

\author{S.V.~Streltsov}
\affiliation{Institute of Metal Physics, Russian Academy of Sciences,
620041 Yekaterinburg GSP-170, Russia}

\author{S.L.~Skornyakov}
\affiliation{Theoretical Physics and Applied Mathematic Department,
Urals State Thechnical University,
620002, Mira st. 19, Yekaterinburg, Russia}

\author{D.M.~Korotin}
\affiliation{Institute of Metal Physics, Russian Academy of Sciences,
620041 Yekaterinburg GSP-170, Russia}

\author{V.I.~Anisimov}
\affiliation{Institute of Metal Physics, Russian Academy of Sciences,
620041 Yekaterinburg GSP-170, Russia}

\date{\today}

\begin{abstract}   Effects of Coulomb correlation on LaOFeAs  electronic structure have been investigated by LDA+DMFT(QMC) method. The calculation results show that LaOFeAs is in the regime of intermediate correlation strength with significant part of the spectral density moved from the Fermi energy to Hubbard bands. However the system is not on the edge of metal insulator-transition because increase of the Coulomb interaction parameter value from $U$=4.0 eV to $U$=5.0 eV did not result in insulator state. Correlations affect different d-orbitals not in the same way. $t_{2g}$ states ($xz,yz$ and $x^2-y^2$ orbitals) have higher energy due to crystal filed splitting and are nearly half-filled. Their spectral functions have pseudogap with Fermi energy position on the higher sub-band slope. Lower energy $e_g$ set of d-orbitals ($3z^2-r^2$ and $xy$) have significantly larger occupancy values with typically metallic spectral functions.
\end{abstract}


\pacs{ }
\maketitle

{\it Introduction.}
The very recent report of superconductivity with the
remarkable $T_c$ of 26K in LaO$_{1-x}$F$_x$FeAs \cite{exp} has stimulated
an intense experimental and theoretical activity. The first attempts to calculate electron-phonon coupling in this material \cite{boeri} have shown that it can not give so large superconducting transition temperature $T_c$. That resembles the situation with high-$T_c$ cuprates and rise a question of the strength for Coulomb correlation effects in this material. Recently Haule {\it et al} \cite{haule} have reported results of LDA+DMFT calculations for LaOFeAs with effective impurity model solved by continuous time quantum Monte-Carlo method \cite{CT-QMC}. Their conclusion is that this system is in strongly correlated regime on the edge of metal-insulator transition with only 20-30\% of spectral weight in quasiparticle band. In the present work we report results of LDA+DMFT(QMC) calculations  for LaOFeAs with efective impurity model solved by standard quantum Monte-Carlo method in Hirsch-Fye algorithm \cite{HF86}. Our results confirm that correlation effects in LaOFeAs indeed have sizable effect on electronic structure. However the system is not on the edge of metal-insulator transition. While Haule {\it et al} \cite{haule} claim that Coulomb parameter $U$ increasing on only 0.5 eV from $U$=4 eV value drives the system to insulating state, in our calculations even for $U$=5.0 eV the solution is still clearly metallic.

\begin{figure}
\centering
\vspace{10mm}
\includegraphics[width=0.8\linewidth]{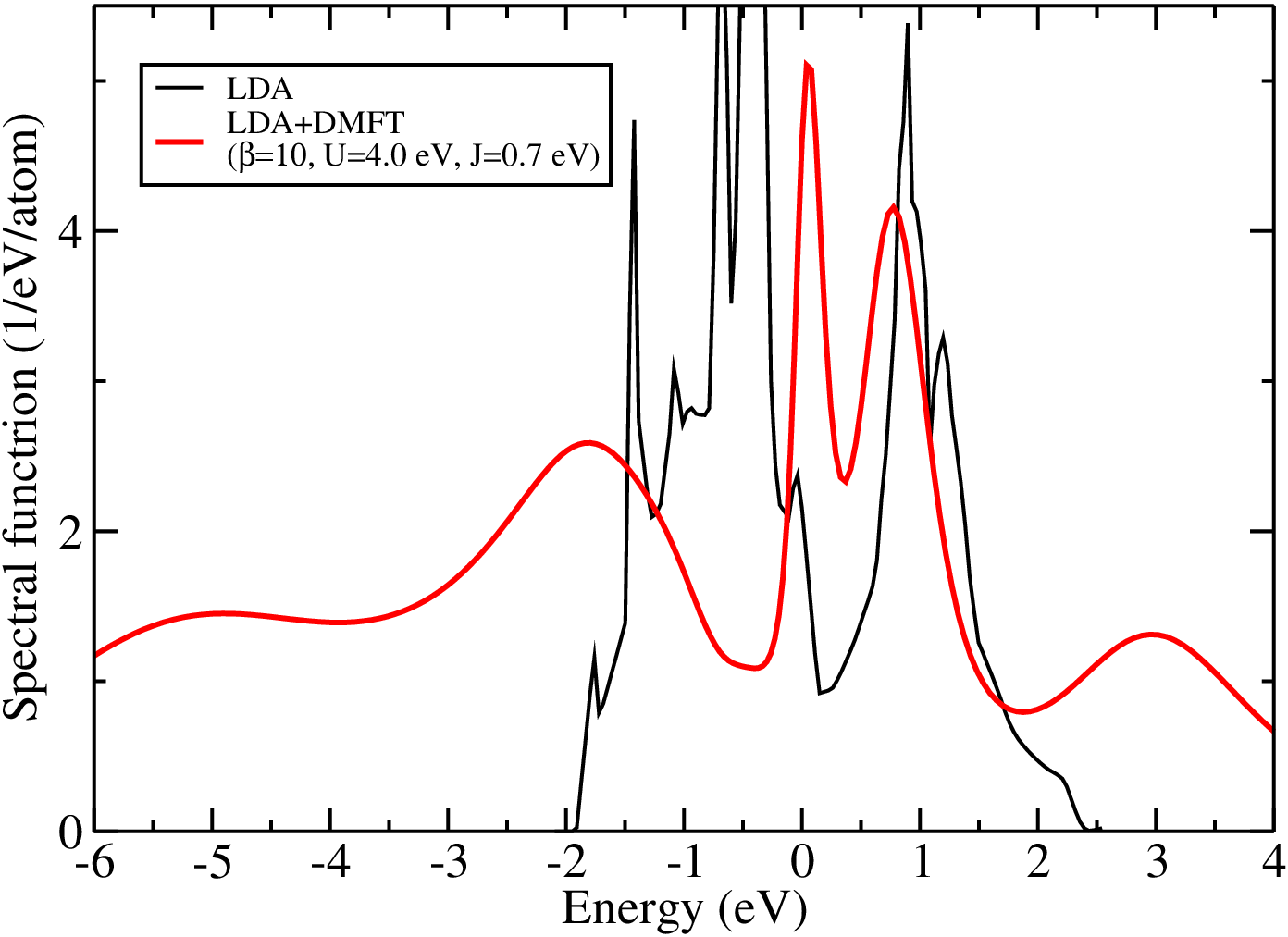}
\caption{(Color online) DOS obtained within LDA approximation (black line) 
and spectral function from LDA+DMFT for $U$=4.0 eV, $J$=0.7 eV (color line).}
\label{fig:dostot}
\end{figure}

{\it Computation details.}
Band structure of LaOFeAs  was calculated in LDA \cite{LDA} approximation
by TB-LMTO-ASA\cite{Andersen84} method. Crystal structure was taken from ~\cite{exp}. The results agree well with previously reported calculations by FP-LAPW method \cite{singh}. In Fig.~\ref{fig:dostot} density of states for energy region corresponding to the bands formed by Fe3d-orbitals is presented. Fermi level corresponding to 6 electrons per formula unit in d-band is situated near deep minimum between two sub-bands. These  sub-bands have their origin in  bonding-antibonding separation due to strong hybridization between d-states in FeAs layers.

Using eigenvalues and eigenfunctions from LDA calculations for bands formed by Fe3d-orbitals  we have calculated Wannier functions via projection procedure \cite{Marzari97, Anisimov05}. The calculated Hamiltonian matrix ${\hat H}({\bf k})$ in  Wannier functions basis has dimensions 10$\times$10 (5 d-orbitals per Fe ion and two Fe ions in unit cell) and by construction has eigenvalues exactly the same as for ten LDA bands formed by Fe-3d orbitals.

This Hamiltonian was used in LDA+DMFT\cite{LDA+DMFT}  calculations (for detailed description of the present computation scheme see \cite{Anisimov05}). The Coulomb interaction parameter value $U$=4 eV and Hund's parameter $J$=0.7 eV used in our work were the same as in previous LDA+DMFT calculations by Haule and Kotliar \cite{haule}. 
The effective impurity model for DMFT was solved by  QMC method in Hirsh-Fye algorithm \cite{HF86}. Calculations were performed for the value of inverse temperature $\beta$=10 $eV^{-1}$. Inverse temperature interval $0<\tau<\beta$ was divided in 100 slices. 4 millions QMC sweeps were used in self-consistency loop within LDA+DMFT scheme and 12 millions
of QMC sweeps were used to calculate spectral functions.

\begin{figure}
\centering
\vspace{10mm}
\includegraphics[width=0.8\linewidth]{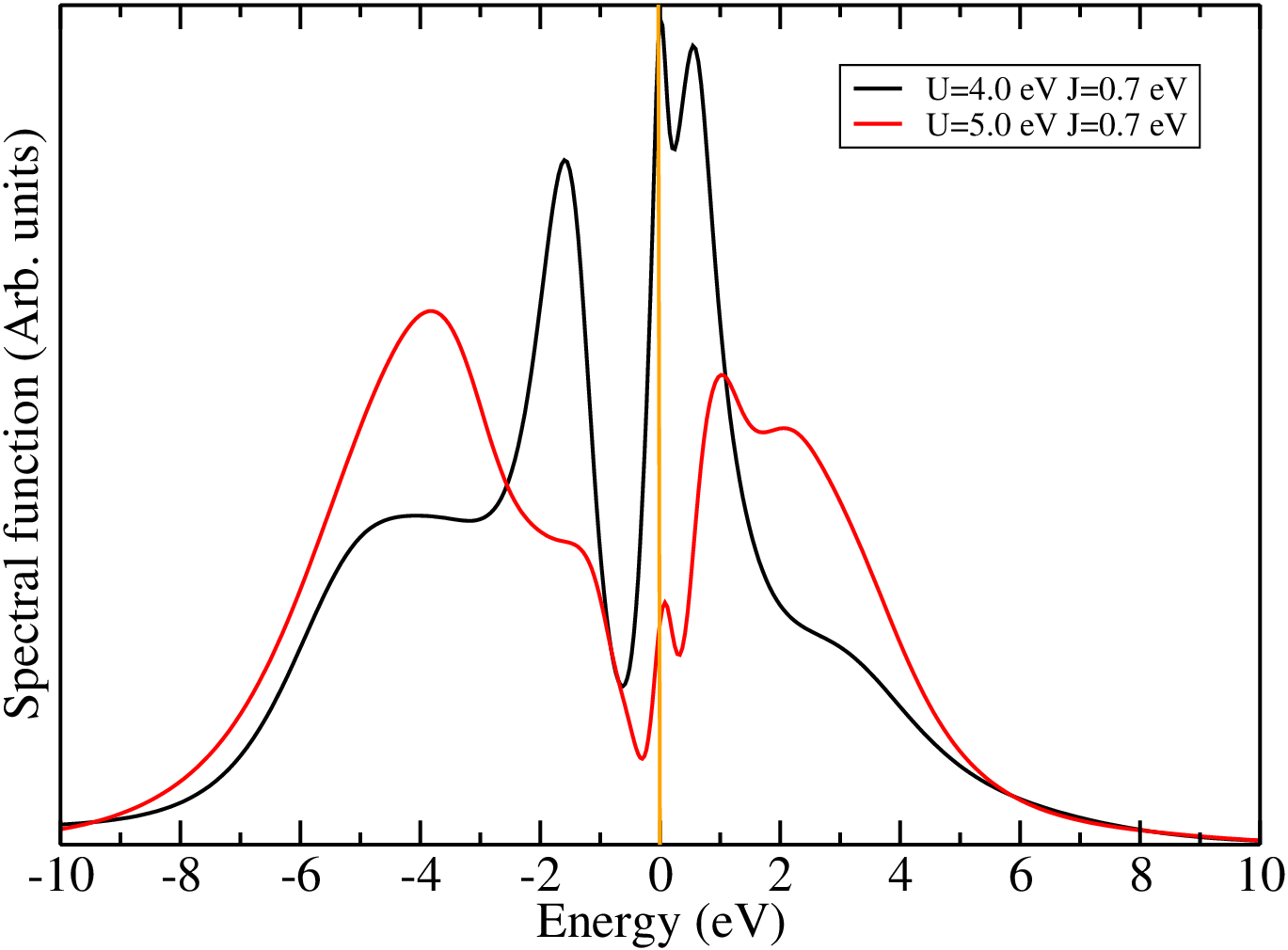}
\caption{(Color online) Spectral function from LDA+DMFT for $U$=4.0 eV (black line) and $U$=5 eV (color line) ($J$=0.7 eV for both).}
\label{fig:dos4-5}
\end{figure}

{\it Results and discussion.}  
Total spectral function calculated in LDA+DMFT in comparison with total density of states (DOS) obtained within LDA are presented in Fig.~\ref{fig:dostot} (see also Fig.~\ref{fig:dos4-5} for wider energy range). Correlation effects are strong enough to move part of the spectral density from Fermi energy to ``shoulders'' at -4 eV and +4eV. However in contrast to Haule {\it et al} \cite{haule} spectra where strong Hubbard band at -4 eV is separated from the weak quasiparticle band by a deep wide depression in (-2 eV, 0 eV) area, in our results a main part of spectral density is still in the interval from -2eV till +2 eV. In this energy region LDA+DMFT spectral function preserves the general shape of LDA DOS with two subbands divided by a deep minimum.  The essential result of correlation is the position of the Fermi energy that has moved from the edge of lower subband in LDA DOS to the top of the upper subband. Exactly on the Fermi energy a sharp peak has developed.

\begin{figure}
\centering
\vspace{10mm}
\includegraphics[width=0.8\linewidth]{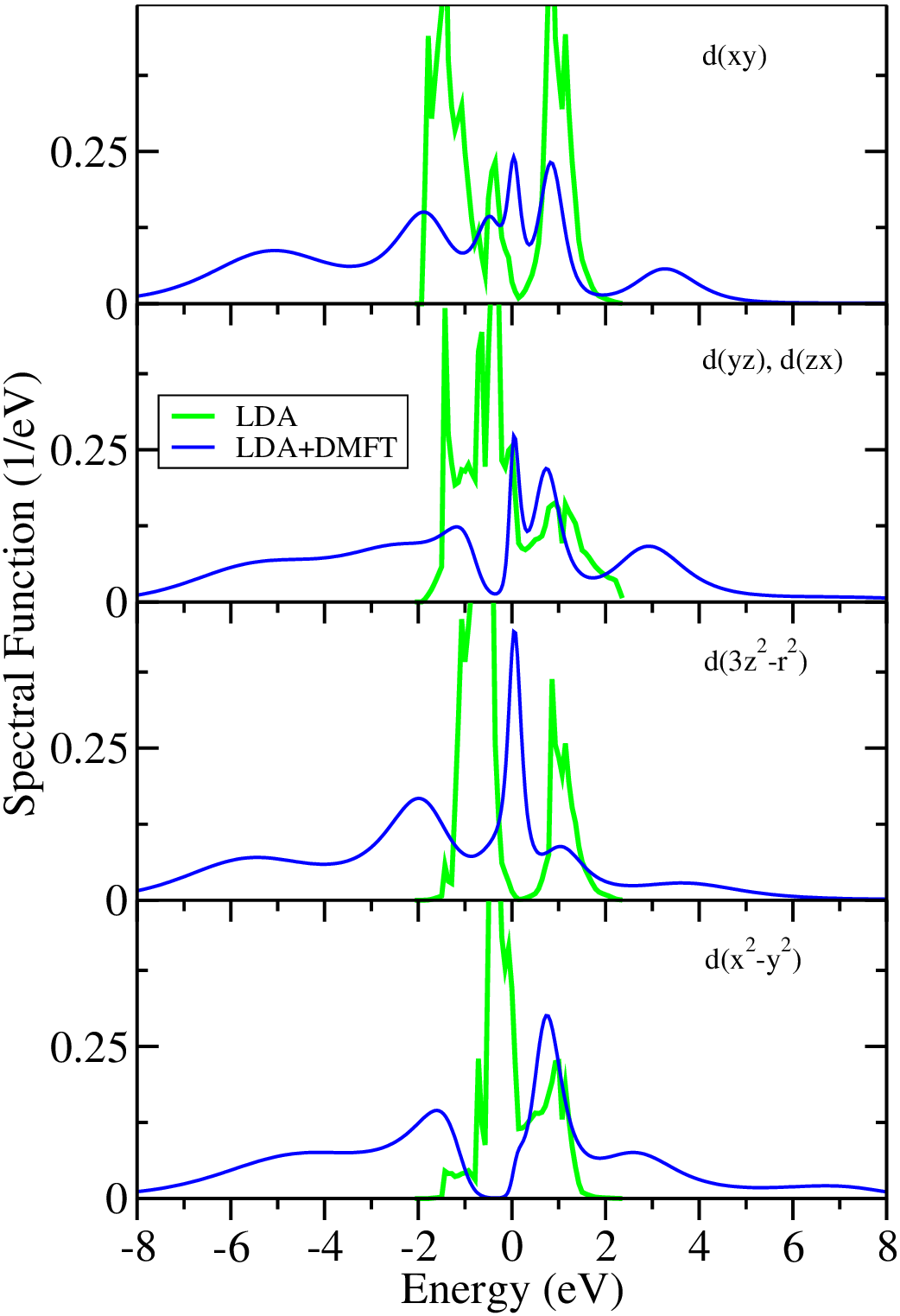}
\caption{(Color online) PDOS obtained within LDA approximation (green lines)
and LDA+DMFT partial spectral functions for $U$=4.0 eV, $J$=0.7 eV (blue lines).}
\label{fig:pdos}
\end{figure}

LaOFeAs crystal structure has tetragonal symmetry and hence Green function for d-orbitals is diagonal with different matrix elements for every one of $3z^2-r^2$, $x^2-y^2$, $xy$ orbitals and doubly degenerate elements for $xz,yz$ set of orbitals. Fe ion has tetrahedral coordination of four As ions with slight tetragonal distortion of the tetrahedron. In tetrahedral symmetry group $T_d$ five d-orbitals should be split by crystal field  on low-energy doubly degenerate set $3z^2-r^2,xy$  corresponding to irreducible representation $e_g$ \cite{xy} and high-energy triply degenerate set $x^2-y^2,xz,yz$ for representation $t_{2g}$.  We have calculated Wannier functions energy and have found that $t_{2g}$-$e_g$ crystal field splitting parameter is very small $\Delta_{cf}\approx$0.25 eV. The slight tetragonal distortion of the tetrahedron gives additional  splitting of $t_{2g}$ and $e_g$ levels with the following values for orbital energies (energy of the lowest $xy$ orbital is taken as a zero): $\varepsilon_{xy}$=0.0 eV, $\varepsilon_{3z^2-r^2}$=0.03 eV, $\varepsilon_{xz,yz}$=0.26 eV, $\varepsilon_{x^2-y^2}$=0.41 eV.
Intersite hybridization between d-orbitals is much stronger than  crystal field splitting (the total d-band width is $\approx$4 eV). As a result all five d-orbitals form common d-band with sub-bands originated not from crystal filed splitting, as it is the case in octahedral coordinated oxides, but due to strong bonding-antibonding separation (see Fig.~\ref{fig:pdos}). $xy$-orbital  that has lobes directed along Fe-Fe bond in FeAs planes shows specially strong bonding-antibonding separation.

In Fig.~\ref{fig:pdos} we present orbitally resolved spectral functions obtained in LDA+DMFT calculation with $U$=4.0 eV. Comparison with corresponding LDA partial DOS shows that different orbitals are affected by correlations not in the same way. $x^2-y^2$ curve has nearly insulator shape with a pronounced pseudogap below the Fermi energy. Another $t_{2g}$ orbitals $xz,yz$ also show presence of peudogap but Fermi energy is on the slope of upper sub-band, with a general form typical to slightly electron doped Mott insulator. $e_g$ set of d-orbitals ($3z^2-r^2$ and $xy$) show clearly metallic behaviour with well developed quasiparticle bands and no pseudogap. 

Occupation numbers obtained from  LDA+DMFT calculations have the following values:
$n_{xy}$=0.672, 
$n_{yz}$,$n_{zx}$=0.565, 
$n_{3z^2-r^2}$=0.686 and 
$n_{x^2-y^2}$=0.512. 
These values mean that one of $t_{2g}$ states, $x^2-y^2$ orbital that has highest crystal field energy ($\varepsilon_{x^2-y^2}$=0.41 eV) is nearly half-filled. Energy of the other of $t_{2g}$ states ($xz,yz$-orbitals)  is 0.15 eV lower ($\varepsilon_{xz,yz}$=0.26 eV) and these orbitals are slightly more occupied.  $e_g$ set of d-orbitals ($3z^2-r^2$ and $xy$) which  have lowest energy due to crystal field splitting ($\varepsilon_{xy}$=0, $\varepsilon_{3z^2-r^2}$=0.03 eV) have significantly larger occupancy. This fact agrees with our analysis of 
 orbitally resolved spectral functions in Fig.~\ref{fig:pdos}. Half-filling  of $x^2-y^2$ orbital ($n_{x^2-y^2}$=0.512) corresponds to Mott insulator shape of the spectral function. Deviation from half-filling for $xz,yz$ orbitals ($n_{yz}$,$n_{zx}$=0.565) results in light electron doped Mott insulator. The occupancies of  $e_g$ orbitals are much larger than $t_{2g}$ values and corresponding spectral functions are characteristic for correlated metal.

Authors of \cite{haule} report that ``slightly enhanced
Coulomb repulsion ($U$= 4.5 eV) opens the gap'' so that the system is on the edge of metal-insulator transition. We have repeated LDA+DMFT(QMC) calculations with even large Coulomb parameter $U$=5.0 eV (Hund's parameter $J$=0.7 eV). The resulting total spectral function is presented in  Fig.~\ref{fig:dos4-5} in comparison with result for $U$=4.0 eV. While increasing of  $U$ value leads to redistribution of spectral density from the peaks around Fermi to lower and upper Hubbard bands, the system still shows clearly metallic character with sharp peak on the Fermi energy. One can conclude that for $U$=4.0 eV value the system is not close to metal-insulator transition.

\begin{figure}
\centering
\vspace{15mm}
\includegraphics[width=0.8\linewidth]{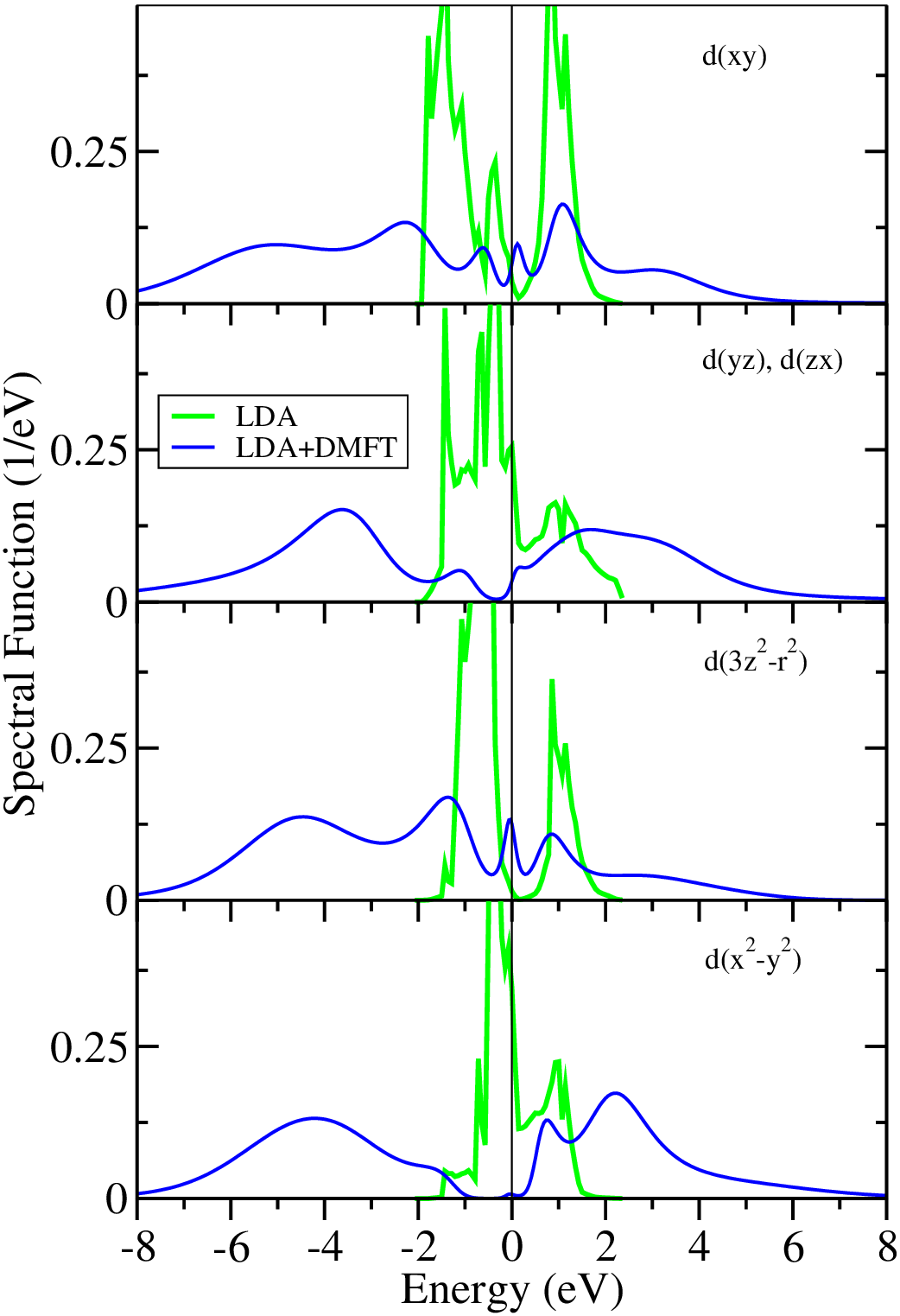}
\caption{(Color online) PDOS obtained within LDA approximation (green lines)
and LDA+DMFT partial spectral functions for $U$=5.0 eV, $J$=0.7 eV (blue lines).}
\label{fig:pdos5}
\end{figure}

It is interesting to analyze  evolution of orbitally resolved spectral functions  with increasing $U$ value (Fig.~\ref{fig:pdos5}). For $U$=5 eV $t_{2g}$ states comparing with $U$=4 eV results (Fig.~\ref{fig:pdos}) become fully insulating with quasiparticle peak absent for $xz,yz$ orbitals and Fermi level position inside the gap for $x^2-y^2$ spectral function. However $e_g$ set of d-orbitals ($3z^2-r^2$ and $xy$) still have a strong quasiparticle peak on the Fermi energy. That type of electronic structure with partial localization resembles ``Orbitally Selective Mott transition'' \cite{OSM} effect that was proposed for Ca$_{2-x}$Sr$_x$RuO$_4$ materials. Mott transition for $x^2-y^2$ spectral function is not 100\% pure because one can observe weak peak on the Fermi energy inside the gap. This peak can be due to the mixing of $x^2-y^2$ orbitals with metallic $e_g$ states. The evolution of orbital occupancies with increasing $U$ value agrees with better localization of $t_{2g}$ states for $U$=5 eV: for $(xz,yz),x^2-y^2$ orbitals occupancy values have decreased from (0.565, 0.512) to (0.535, 0.505) and for $e_g$ set of d-orbitals ($3z^2-r^2$ and $xy$) the occupancies have increased from (0.686, 0.672) to (0.750, 0.672).

We have found that inclusion of Hund parameter $J$=0.7 eV into calculations dramatically changes the effect of correlations on  LaOFeAs. LDA+DMFT(QMC) calculations with $U$=4eV and $J$=0 eV results in very weakly correlated electronic structure with quasiparticle renormalization amplitude $Z\approx$0.8. This agrees with the results of  \cite{cox}, where effect of d-band degeneracy $N_d$ on critical values for $U/W$ ratio needed for metal-insulator transition was investigated. The authors of  \cite{cox} have estimated the critical $U$ value as $U_c\approx\sqrt{N_d}U_c^0-N_d J$, where $U_c^0$ is critical $U$ value for non-degenerate Hubbard model. With LDA band width value  $W\approx$ 4 eV and Coulomb interaction parameter $U$= 4.0 eV  their ratio $U/W$ is close to 1. With $J$=0 orbital degeneracy $N_d$=5  would lead to critical $U$ value increased by a factor of $\sqrt{5}$ or effective decrease of $U/W$ ratio more than twice thus resulting in weakly correlated regime. However $J$=0.7 eV will compensate increase of $U_c$ due to the $\sqrt{5}$ factor and $U/W\approx$1 should result in intermediate correlation strength regime.

{\it In conclusion.}
The LDA+DMFT(QMC) calculations show that LaOFeAs is in intermediate correlation strength regime. Hubbard bands in spectral function are well pronounced but still a major part of spectral density is in the bands around Fermi energy. The system is not close to metal-insulator transition because even increase of Coulomb interaction parameter value on 1 eV does not lead to insulating state.

{\it Acknowledgments.}
Authors thank Jan Kune\v{s} for his DMFT(QMC) computer code used in our calculations. Support by the
Russian Foundation for Basic Research under Grant No.
RFFI-07-02-00041, Civil Research and Development Foundation together with
Russian Ministry of science and education through program Y4-P-05-15,
Russian president grant for young scientists MK-1184.2007.2 and
Dynasty Foundation
 is
gratefully acknowledged.


\begin{thebibliography}{99}
\bibitem{xy}
The coordinate axes $x,y$ in LaOFeAs crystal structure are rotated on 45 degrees from the standard tetrahedral notation so that $xy$ and $x^2-y^2$ orbitals are interchanged.
\end{thebibliography}
\end{document}